\documentclass{article}[draft]

\usepackage{booktabs}       
\usepackage{arxiv}
\usepackage{float}
\usepackage[usenames,dvipsnames]{color}
\usepackage{placeins}
\usepackage{multirow}

\RequirePackage{uclthesis}

\usepackage[sorting=none,style=numeric-comp]{biblatex}
\renewbibmacro*{in:}{%
  \iffieldundef{journaltitle}
    {}
    {\printtext{\bibstring{in}\intitlepunct}}}

\newcommand{\printtitle}{Transformers for Charged Particle Reconstruction in High Energy Physics}

\title{\printtitle}

\author{
Samuel Van Stroud\textsuperscript{1},
Philippa Duckett\textsuperscript{1},
Max Hart\textsuperscript{1},
Nikita Pond\textsuperscript{1},\\
Sébastien Rettie\textsuperscript{1,2},
Gabriel Facini\textsuperscript{1},
and Tim Scanlon\textsuperscript{1}
\\
\textsuperscript{1}\textit{Centre for Data Intensive Science and Industry, University College London},
\textsuperscript{2}\textit{CERN}
}

\renewcommand{\shorttitle}{\printtitle}

\hypersetup{
    colorlinks,
    pdftitle={Transformers for Charged Particle Track Reconstruction in High Energy Physics},
    pdfsubject={},
    pdfauthor={Samuel Van Stroud et al},
    pdfkeywords={Particle Physics, ML Reconstruction, Object Detection},
    citecolor=blue,
    urlcolor=blue,
    linkcolor=blue,
}

\begin{document}

\newcommand{\effperf}{\varepsilon^\mathrm{Perfect}}
\newcommand{\effperfpnine}{$\varepsilon^\mathrm{Perfect}_{\geq 0.9}$}

\maketitle

\vspace{8em}
\begin{abstract}
Charged particle reconstruction, the identification and characterisation of particles from collision data, is fundamental to nearly all research at particle colliders like the Large Hadron Collider (LHC). With the High-Luminosity upgrade (HL-LHC), particle multiplicities will increase substantially, overwhelming traditional track reconstruction algorithms and presenting computational bottlenecks. Here, we introduce a proof-of-concept for a powerful new method for charged particle reconstruction inspired by state-of-the-art machine learning (ML) approaches in computer vision. Our model leverages Transformer neural networks to efficiently filter relevant signals and fully reconstruct particle trajectories, directly tackling the computational complexity that traditional methods face. Evaluated on the widely used TrackML dataset, our approach achieves state-of-the-art tracking efficiency (97\%) and a low fake rate (0.7\%), requiring just 97 milliseconds to reconstruct on average 1,300 particle trajectories from 55,000 detector hits for particles with transverse momentum above \SI{750}{\MeV}. These results represent a significant milestone in both performance and speed, demonstrating a shift toward unified, scalable ML solutions that offer substantial improvements for collider experiments.
\end{abstract}
\vfill


\newpage
\clearpage
\section{Introduction}
\label{sec:intro}
Modern particle colliders, such as the Large Hadron Collider (LHC)~\cite{LHC} at CERN, explore fundamental physics by producing vast numbers of collision events, each containing thousands of particles.
Analysing these collisions is essential to addressing open questions in physics, including the properties of the Higgs boson~\cite{HDBS-2018-58, CMS-HIG-23-006}, the nature of dark matter, and the fundamental structure of the universe.
A core task in analysing collision data is reconstructing the trajectories (\textit{tracks}) of charged particles produced in these collisions. Track reconstruction provides particle momenta and charge measurements which are critical for nearly all downstream tasks, including lepton reconstruction~\cite{EGAM-2021-01, MUON-2018-03, CMS-EGM-17-001, CMS-MUO-16-001}, hadronic~$\tau$ reconstruction ~\cite{atlas_tau_reco, CMS-TAU-16-003}, particle flow algorithms~\cite{PERF-2015-09,CMS-PRF-14-001}, and jet flavour identification~\cite{FTAG-2019-07,2025arXiv250519689A,CMS-BTV-16-002}.

The upcoming High-Luminosity upgrade of the LHC (HL-LHC) will dramatically increase event complexity. 
HL-LHC beam crossings occur every \SI{25}{ns}, corresponding to a crossing rate of 40 MHz. 
With each beam crossing producing up to 200 proton-proton collisions, the total collision rate will surpass eight billion collisions per second, roughly a factor of four increase from current LHC operations.
The average number of particles produced in each event will increase accordingly to $\mathcal{O}(10^4)$.
This surge in particle multiplicity poses a formidable computational challenge for track reconstruction, as traditional algorithms typically scale worse than quadratically with the number of particle-detector interactions (\textit{hits}) per event.
Consequently, they rapidly become computationally prohibitive at HL-LHC multiplicities~\cite{atlas_hllhc_comp, cms_hllhc_comp}.
Addressing this computational bottleneck is a critical challenge, as efficient and accurate particle tracking directly impacts the physics discovery potential at the HL-LHC.


In this work, we address these computational challenges by developing a new charged particle tracking method built upon recent innovations in machine learning.
Our approach introduces two novel ideas: a Transformer-based~\cite{VaswaniAttention} \textit{hit filtering} stage and a MaskFormer-inspired~\cite{maskformer} \textit{track reconstruction} stage. 
The hit filtering stage uses windowed self-attention, originally developed for language processing tasks, to efficiently reduce the initial hit multiplicity. 
This step avoids the costly combinatorial processing and complex graph construction inherent in prior methods.  
The subsequent track reconstruction stage leverages a MaskFormer architecture, a type of Transformer originally designed for image segmentation, to simultaneously identify tracks, assign hits, and estimate their physical properties, removing the need for separate off-model post-processing steps.
Recent success applying Transformer-based models to vertex reconstruction~\cite{van2023vertex} further demonstrates the architecture's potential for general particle physics reconstruction tasks.

Evaluated on the widely used TrackML dataset (see \cref{sec:dataset}), our method achieves state-of-the-art tracking efficiency (97\%) with low fake rates (0.6\%), and inference times of approximately \SI{100}{ms} per event. 
This represents a substantial improvement compared to typical inference times of seconds to tens of seconds per event observed with traditional CPU-based tracking algorithms~\cite{amrouche2023tracking}.
However, further development is required before operational deployment.
These steps include validation under realistic detector conditions, integration within experimental software frameworks, and evaluation of computational resources in production scenarios.

Beyond general-purpose collider experiments, specialised LHC experiments like LHCb~\cite{LHCb}, ALICE~\cite{ALICE}, and experiments beyond the LHC such as Mu3e~\cite{mu3e} and LUXE~\cite{luxe}, also rely on precise and scalable tracking solutions. 
Our fully-learned approach, inherently flexible and adaptable, has potential as a general solution across diverse experimental setups.
Integrating it into experiment-independent track reconstruction toolkits, such as A Common Tracking Software (ACTS)~\cite{ACTS}, would significantly streamline efforts to develop efficient tracking for new particle detectors.

The structure of this article is as follows. In \cref{sec:related-work}, we present an overview of related research. \cref{sec:model} introduces our Transformer-based model architecture, detailing both the hit filtering and MaskFormer track reconstruction components.
\cref{sec:experiments} includes a description of the TrackML dataset, training procedures, and experiments used. The results are discussed in \cref{sec:results}, demonstrating the scalability and effectiveness of our approach compared to existing methods. 
Finally, \cref{sec:conclusion} includes a summary of our contributions and discusses future directions and broader implications of this work for high-energy physics.

\section{Related Work}
\label{sec:related-work}

Trajectories of charged particles are observed through their non-destructive interaction with sensitive detector elements in specialised silicon tracking detectors.
In general purpose collider experiments, these detectors are typically arranged around the collision region in concentric cylindrical (barrel) layers, with disks (endcaps) at each end and a magnetic field oriented along the beam-line.
Due to the magnetic field, charged particles follow helical paths, curving in the transverse plane resulting in characteristic patterns of discrete 3-dimensional position measurements, referred to as \textit{hits}, across detector layers.
Building upon this fundamental information, a variety of approaches to track reconstruction have been developed to infer the original particle trajectories from hit data. 
This section will review relevant examples of these methods.

\subsection{Traditional Track Reconstruction Methods}
Charged particle tracking has historically relied on computational methods designed to handle detector data efficiently.
While these algorithms have been successful in existing collider experiments, they typically scale poorly (worse than quadratically) with increasing particle multiplicities. 
This scaling behaviour presents critical computational challenges as collision densities dramatically increase with the HL-LHC upgrade~\cite{atlas_hllhc_comp, cms_hllhc_comp}.
A summary of traditional solutions follows.

\vspace{-0.5em}
\paragraph{Combinatorial Kalman Filter (CKF)}
CKF tracking algorithms have been the standard approach for particle track reconstruction since their introduction to high-energy physics~\cite{Fruhwirth:1987fm}, and their widespread adoption in experiments at the Large Electron-Positron (LEP) collider~\cite{ALEPH,DELPHI}. 
These algorithms remain the dominant tracking method used by current collider experiments, including those at the LHC~\cite{IDTR-2022-04, CMS-TRK-11-001, LHCb-Tracking, ALICE-Tracking}. 
Both ATLAS and CMS employ CKFs techniques, which begin with computationally intensive track seeding~\cite{IDTR-2022-04, CMS-TRK-11-001}. 
Track seeding typically involves combinational algorithms that identify initial track candidates (seeds) from small groups of detector hits chosen based on geometric and kinematic constraints. 
Following seeding, tracks are built by iteratively adding hits and refining estimates of track parameters.
Resolving competition between multiple track candidates for shared hits often requires further processing. 
While CKF tracking achieves high accuracy and robustness under realistic detector conditions, it becomes prohibitively resource-intensive for HL-LHC environments~\cite{atlas_hllhc_comp, cms_hllhc_comp}. 
Despite algorithmic optimisations~\cite{ATL-PHYS-PUB-2019-041}, significant computational challenges remain: at HL-LHC multiplicities CKFs can take $\mathcal{O}(\SI{5}{\s})$ per event on a single CPU core to reconstruct charged particles with a transverse momentum (\pT) greater than \SI{1}{GeV}~\cite{ATLAS-PhaseII-EFTrk,cms_phase2_trig,amrouche2023tracking}. 
More recent developments by ATLAS demonstrated a sub~\SI{2}{\s} configuration using ACTS in a ``fast'' mode, optimised for the data selection system (trigger)~\cite{IDTR-2025-04}.

\vspace{-0.5em}
\paragraph{Cellular Automata (CA)}
CA~\cite{cellular-automata} approaches mitigate combinatorial complexity by incrementally linking nearby hits into track segments based on local geometric criteria. 
It has been used in a variety of completed experiments~\cite{CHEP1992,Glazov1993,Bussa1996,Kisel1997}.
CA-based approaches offer strong scalability and are already deployed in production GPU-accelerated tracking pipelines, such as CMS’s Patatrack system in the trigger~\cite{Bocci:2020pmi,CMS-DP-2023-028}. 
The CMS experiment has also integrated CA-based track seeding into its Phase-II pixel detector upgrade plans~\cite{CMS-TDR-Tracker}.
ALICE employs a GPU-accelerated CA method to reconstruct tracks in heavy-ion collisions, leveraging CA's ability to be parallelised and computational efficiency~\cite{ALICE-GPU-tracking}. 
However, cellular automata require careful tuning of geometric rules and thresholds, and their performance can degrade significantly in dense particle environments where track overlap and ambiguous hit association rates increase.

\vspace{-0.5em}
\paragraph{Hough Transform}
Hough transform methods~\cite{hough-transform, ATLAS-PhaseII-EFTrk} reconstruct tracks by transforming hit positions into parameter space (e.g., curvature and angle), where collinear or curved track hits appear as peaks.
Tracks are then identified by detecting these peaks. 
Hough transforms can efficiently manage combinational complexity and identify tracks in noisy environments with robustness against missing hits and noise from its global pattern-recognition approach.
However, within higher track multiplicity settings, the computational demands and susceptibility to spurious peak formation from coincidental hit alignments require complex multi-stage solutions.

\subsection{Emerging Solutions}
The computational demands of real-time track reconstruction at facilities like the HL-LHC have spurred exploration into hardware accelerators.
FPGAs and GPUs are prominent candidates due to their inherent parallel-processing capabilities.
Both CMS and ATLAS are actively developing FPGA-based track finders for their HL-LHC trigger upgrades, with studies indicating potential for sub-microsecond latencies~\cite{CMS-TDR-Tracker,ATLAS-PhaseII-EFTrk}.
GPUs have also proven effective, accelerating software-level tracking for experiments like ALICE and LHCb by leveraging massive parallelism for tasks such as Cellular Automata and Kalman filtering~\cite{ALICE-GPU-tracking,LHCb-Allen}.

Beyond traditional algorithms, GPUs are increasingly facilitating advanced Machine Learning (ML)-based tracking methods.
This section reviews several modern approaches to track reconstruction, many of which target GPU deployment.

\vspace{-0.5em}
\paragraph{The TrackML Challenge}
The TrackML challenge~\cite{amrouche2020tracking,amrouche2023tracking} catalysed community engagement in developing efficient and scalable track reconstruction algorithms.
The dataset (described \cref{sec:experiments}) comprises simulated HL-LHC collision events under idealised detector conditions, with solutions ranked by accuracy (weighted fraction of correctly assigned hits) and speed (time per event on two CPU cores).
The winning submission of the throughput phase, "Mikado"~\cite{amrouche2020tracking}, employed 60 passes of combinational searches with localised helix fitting and optimised data structures to achieve 94.4\% accuracy at \SI{0.56}{\s/event}.
While innovative, Mikado's reliance on complex, hand-crafted logic and an extensive, largely ad hoc optimisation process for its $\sim$20k tunable parameters limits its wider applicability.

\vspace{-0.5em}
\paragraph{GNN4ITk}
Graph Neural Networks (GNNs) \cite{gnn_original_paper} have emerged as a leading approach to address the computational challenges of track reconstruction at the HL-LHC.
The ATLAS collaboration's GNN4ITk pipeline \cite{caillou2022atlas,caillou2024physics} (based on \cite{Ju:2021ayy,Biscarat:2021dlj}) involves graph construction from hits, GNN-based edge scoring, edge filtering, and iterative graph segmentation for track candidate extraction.
While achieving high tracking efficiency (over 90-99\% depending on purity criteria, with ~0.1\% fake rate \cite{Caillou:2023gnn}), the initial graph construction remains a bottleneck, even with recent optimizations, processing remains in the 500-\SI{700}{\ms} range for events with $\mathcal{O}(10^5)$ hits~\cite{ATL-PHYS-PUB-2024-018}.
Furthermore, the reliance on custom pre-processing (graph construction) and post-processing (graph segmentation) marries a learned GNN-based edge filtering step with non-learned components, which may not be optimal compared to a fully end-to-end learned approach.

\vspace{-0.5em}
\paragraph{HGNN}
Hierarchical Graph Neural Networks (HGNN) \cite{liu2023hierarchical} offer an alternative to GNN4ITk's iterative segmentation by grouping nodes into \textit{super-nodes}, enlarging the GNN's receptive field. Though HGNN can improve reconstruction efficiency, its reported high fake rate (over 50\%) and increased computational cost necessitate further post-processing, limiting its feasibility for HL-LHC deployment.

\vspace{-0.5em}
\paragraph{Object Condensation}
Object condensation \cite{lieret2024object} reconstructs tracks by clustering hits in a learned latent space.
Representative hits are selected to characterise tracks, with remaining hits assigned via an off-model clustering algorithm.
OC also employs an edge classification step to prune hit connections.
A potential limitation is its reliance on single representative hits, which couples hit feature extraction and object identification. 
Furthermore, the current clustering-based hit assignment is not learned end-to-end and does not natively support assigning hits to multiple tracks, a scenario commonly encountered in complex collision events with closely spaced or overlapping particle trajectories.

\vspace{-0.5em}
\paragraph{Experimental Transformer approaches}
Early explorations of Transformers for track reconstruction \cite{2024arXiv240707179C,2024arXiv240210239H,2024arXiv240721290M,miao2024localitysensitivehashingbasedefficientpoint} have typically focused on direct hit classification or sequential modeling of tracks with autoregressive models.
Detailed comparisons are often challenging as these works may not have been validated on the full-scale TrackML dataset or may not fully characterise performance across tracking efficiency, fake rate, and timing.

\vspace{-0.5em}
\paragraph{MaskFormers}
Recent computer vision advancements, including bounding box prediction \cite{2015arXiv150602640R,2020arXiv200512872C} and sophisticated instance segmentation \cite{2017arXiv170306870H,maskformer}, offer promising alternatives to geometric deep learning for ML-based track reconstruction.
When analysing objects depicted within images, the MaskFormer architecture \cite{maskformer,mask2former} effectively identifies multiple instances of the same object class.
It achieves this by assigning pixels to distinct object instances, each represented by a unique segmentation mask.
By replacing the image pixels with an unordered set of hits and the image-depicted objects with charged particle tracks, we leverage MaskFormers to address the challenges of track reconstruction.
The utility of MaskFormers in HEP was recently demonstrated in \rcite{van2023vertex}, which successfully adapted the \maskformer architecture to reconstruct displaced secondary decay vertices.
In this work, we apply the same architectural foundation to the distinct challenge of track reconstruction.
The successful application of a common architecture to two different reconstruction tasks (vertex finding and track reconstruction) suggests a promising path towards more unified and less task-specific approaches in particle physics reconstruction.

\cref{tab:ml-comparison} summarises the key aspects of the various ML-based approaches to charged particle track reconstruction discussed in this section, including a comparison of the minimum transverse momentum (\ptmin), maximum pseudorapidity ($\eta$) used to define target particles (for more information see \cref{sec:experiments}).

\begin{table}[h]
    \centering
    \begin{tabular}{lccccc}
        \toprule
        \midrule
        & \ptmin & \etamax & Layers Used & Preprocessing & Postprocessing\\
        \midrule
        GNN4ITk \cite{caillou2024physics} & \SI{1}{\GeV} & 4.0 & Pixel + strip & Edge classification & Graph traversal \\
        HGNN \cite{liu2023hierarchical} & \SI{1}{\GeV} & 4.0 & Pixel + strip & Edge classification & GMPool \\
        OC \cite{lieret2024object} & \SI{900}{\MeV} & 4.0 & Pixel & Edge classification & Clustering \\
        \midrule
        \multirow{3}{*}{This Work} & \SI{600}{\MeV} & 2.5 & Pixel & Hit filtering & None \\
        & \SI{750}{\MeV} & 2.5 & Pixel & Hit filtering & None \\
        & \SI{1}{\GeV} & 4.0 & Pixel & Hit filtering & None \\
        \midrule
        \bottomrule
    \end{tabular}
    \vspace{1em}
    \caption{
        Comparison of ML-based approaches to charged particle track reconstruction.
        Tracking detectors at general purpose collider experiments are composed of pixel and strip layers of silicon-based non-destructive charged-particle detector elements, see \rcite{2019EPJWC.21406037K} for more information.
        More information about the TrackML detector and dataset can be found in \cref{sec:dataset}.
        Our approach is tested with two different limits on \etamax -- 2.5 for the \SI{600}{\MeV} and \SI{750}{\MeV} trainings and and 4.0 for the \SI{1}{\GeV} training (see \cref{sec:train-setup} for more information).
    }
    \label{tab:ml-comparison}
\end{table}

\section{Model Architecture}
\label{sec:model}

\subsection{Leveraging Transformers}\label{sec:Transformers}

Transformers~\cite{VaswaniAttention} have revolutionised natural language processing and computer vision tasks, offering a scalable and efficient architecture for processing sequential data.
The attention mechanism within enables the network to determine the relevance between different input features. 
Each feature (e.g., a detector hit or track) is projected into three learned representations: a \textit{key} which defines what each element offers for comparison, a \textit{value} containing the information to be aggregated, and a \textit{query} which initiates the search for relevant information. 
Attention is computed by taking the dot product of each query with all keys, producing similarity scores that determine how much information to gather from each value.
In our model, queries correspond to latent (i.e., learned) track representations, and the keys and values are derived from the embedded features of detector hits. 
The output of this attention process is a mask for each query, representing the model's confidence that individual inputs (e.g., hits) belong to that query (e.g., a reconstructed track).
In \textit{self-attention} (SA) the queries, keys, and values all come from the same input set, allowing elements (e.g., hits or latent tracks) to attend to each other.
In \textit{cross-attention} (CA) queries from one source (e.g., latent track representations) and keys/values from another (e.g., detector hits), enabling the model to associate tracks with relevant hits.
However, due to Transformer's quadratic complexity in the number of input tokens, custom GNN-based architectures have so far been preferred for particle tracking tasks \cite{caillou2024physics,lieret2024object}.

To neutralise the quadratic complexity of Transformers, we hypothesise that hits only need to attend to nearby hits in the azimuthal angle $\phi$, and apply this powerful prior of $\phi$-locality by ordering hits in $\phi$ and applying sliding window attention \cite{2020arXiv200405150B} with a window size $w$.
This results in $\mathcal{O}(M \times w)$ complexity scaling, which is linear in the number of input hits $M$.
To allow hits to communicate around the $\pm\pi$ boundary, the first $w/2$ hits are appended to the end of the sequence and vice versa.
We further benefit from the fused attention kernels provided by \texttt{FlashAttention2} \cite{2023arXiv230708691D} for efficient computation, and \texttt{SwiGLU} activation \cite{2020arXiv200205202S} for improved performance.

This approach allows us to efficiently encode 60k pixel hits in \SI{25}{\milli\second} on a single GPU, making it suitable for low-latency trigger-level track reconstruction at the HL-LHC.
Scaling to higher hit multiplicities could be achieved through algorithmic optimisations, or simply scaling to multiple GPUs.
A common encoder architecture is used in both the hit filtering and track reconstruction stages of our model, as described in the following sections.
Detailed timing results can be found in \cref{sec:timing}.

\subsection{Hit Filtering Model}

The large number of hits present in each event makes it computationally infeasible to feed all hits directly into the tracking model.
As described in \cref{sec:related-work}, previous approaches to ML-based track reconstruction have required a graph construction stage based on geometric constraints and/or edge classification to reduce the input hit multiplicities.
In contrast, we introduce a Transformer-based hit filtering model to reduce the input hit multiplicities by directly predicting whether each hit is \textit{noise} or \textit{signal}. 
\textit{Noise} hits are defined as hits belonging to particles that we do not wish to reconstruct, for example, particles below a \pt threshold or outside a specified pseudorapidity range, in addition to the intrinsic noise hits which do not belong to any simulation-level particle. The remaining hits are labelled \textit{signal}.

The hit filtering model is composed of an embedding layer, a Transformer encoder, and a dense hit-level classifier.
In the model, the input features of $M$ hits are first passed through an embedding layer to produce a $d=256$-dimensional representation for each hit.
Positional encodings are applied \cite{VaswaniAttention} in the cylindrical hit coordinates (see \cref{sec:dataset}) to allow the SA mechanism to readily identify nearby hits.
In particular, a cyclic positional encoding \cite{Lee2022SpatioTemporalOL} is used for $\phi$.
The initial hit embeddings are then passed into an efficient SA Transformer encoder as described in \cref{sec:Transformers}.
The encoder has twelve layers (eight for the \SI{1}{\GeV} model, see \cref{sec:experiments}), model dimension $d$, and feed-forward dimension $2d$.
A window size of $w=1024$ is used for the sliding window attention.
The output embeddings are classified using a dense network with three hidden layers.

The hit stand-alone filtering step offers broad applicability beyond our \maskformer-based approach to track reconstruction.
It could be integrated as a pre-processing step into other traditional and ML-based tracking pipelines, or repurposed for other tasks such as pileup mitigation.

\subsection{Track Reconstruction Model}

\begin{figure}[htb]
    \centering
    \vspace{-1em}
    \includegraphics[width=0.8\textwidth]{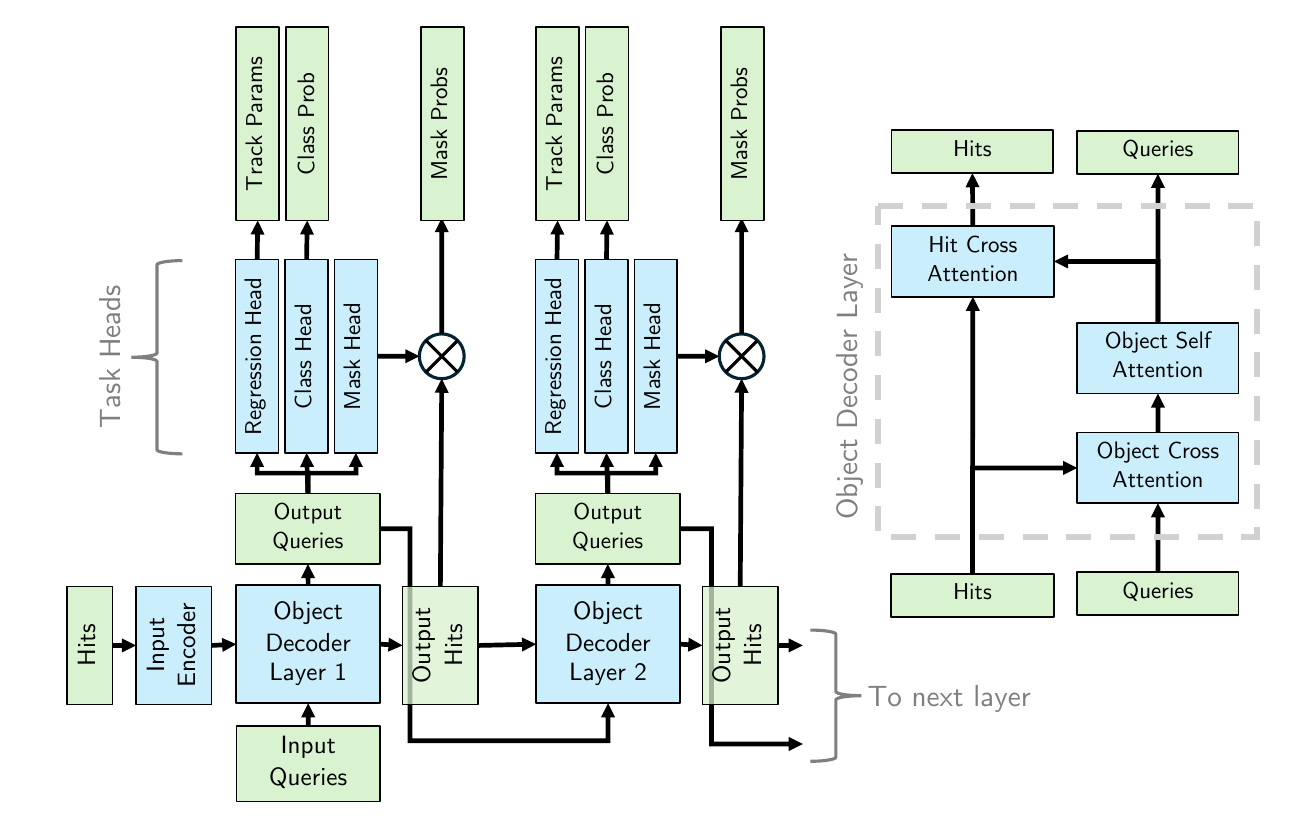}
    \caption{ 
        Overview of the track reconstruction model, with data and operations being shown in green and blue respectively. 
        $M$ input tokens representing the hits are fed into an initial Transformer encoder.
        The object decoder then takes a set of $N$ object queries, which represent tracks, and iteratively updates them with information from the input elements and other object queries.
        Finally, three task heads are used to: categorise each object as being from one of $C+1$ object classes (including a \null class), estimate $R$ regression targets, and predict $N \times M$ binary masks which provide the assignment of input hits to output tracks. The resulting embedded queries and hits are then fed into another decoder layer to refine the predictions and produce an intermediate auxiliary loss for each layer. These decoder layers can be stacked repeatedly to increase accuracy at the expense of computational cost. On the right, a detailed view of an object decoder layer is shown.     
    }
    \label{fig:mf_arch}
\end{figure}

The tracking model follows an encoder-decoder architecture as shown in \cref{fig:mf_arch}, with the encoder processing the input hits and the decoder reconstructing the tracks.
The encoder model is the same as the hit filtering model, except for the window size which is decreased to 512 to compensate for the reduced hit multiplicities after filtering.
The object decoder is a \maskformer-based model \cite{maskformer,mask2former,segmentanything} which forms an explicit latent representation for each track candidate, allowing for joint optimisation of hit assignments and track parameters.
Similar to \rcite{van2023vertex}, we include modifications to handle sparse inputs, and to allow additional regression tasks for the reconstructed tracks.

The object decoder initialises a set of $N$ object queries as learned vectors of dimension $d$.
Each object query represents a possible output track.
The number of object queries $N$ sets the maximum number of tracks that can be reconstructed per event, and is chosen as the maximum number of tracks over the events in the training sample (described in \cref{sec:dataset}).
The object queries are then passed through eight decoder layers.
In each layer, the object queries aggregate information from relevant hits via bi-directional CA, and from other object queries via SA.
The \texttt{MaskAttention} operator \cite{mask2former} is used to generate attention masks from the intermediate mask proposals produced by the preceding decoder layer, encouraging object queries to attend only to relevant hits.
The resulting object queries from the object decoder are processed by three task heads, which predict the track class, track-to-hit assignments, and track properties.
The number of decoder layers and model dimension $d$ was chosen to achieve good performance with reasonable training and inference times. 

Since the number of object queries is fixed, but the number of tracks in each event is variable, a dense binary classifier with a single hidden layer of size $d$ is used to predict whether each object query corresponds to a track or not.
If not, other outputs for that object query are ignored.
The assignment of each of the hits to tracks is given by an $M$-dimensional mask over the input hits for each of the $N$ object queries.
Mask tokens are formed from the query embeddings via a dense network with a single hidden layer of size $2d$.
The mask for each hit is computed by taking the dot product between the updated hit embeddings from the object decoder and the query mask token for each object query.
After applying a sigmoid activation, the mask is binarised, and a value of 1 indicates assignment of the hit to the track, while 0 indicates no assignment.
Using an element-wise sigmoid operation as opposed to applying a softmax over the hits or tracks allows the model to assign a single hit to multiple tracks, which can occur due to random track crossings or from collimated particles in the core of high-\pt jets~\cite{PERF-2015-08}.
Track parameters are regressed using a dense network with four output nodes, with each node corresponding to an output target.
We regress the particle 3-momenta in Cartesian space $(p_x, p_y, p_z)$, along with the $z$ position of the production vertex $v_z$.
The total momentum is not regressed directly, but rather calculated from the component predictions and added to the loss to enforce consistency. Other track parameters, such as the track transverse momentum \pt, angle $\phi$ and pseudorapidity $\eta$, are derived from the regressed Cartesian components of the momentum. 

The training loss is the sum of the loss terms from each of the tasks.
For the object class, a categorical cross-entropy loss $L_\text{CE}$ is used.
For the regression targets, a SmoothL1 loss \cite{2015arXiv150408083G} $L_\text{Regression}$ is used.
The mask loss $L_\text{Mask}$ is a combination of a Dice loss $L_\text{Dice}$ ~\cite{dice} and focal loss $L_\text{Focal}$ ~\cite{focal}.
The total loss is then the weighted sum
\begin{equation}
    L =
    0.1 L_\text{CE}+ 
    \underbrace{   
 2 L_\text{Dice} + 50 L_\text{Focal}}_{L_{\text{Mask}}} +
   0.1  L_{\text{Regression}},
\end{equation}
where the weights are coarsely optimised to provide an good trade-off between the performance of the different tasks.
To ensure that the loss is invariant over permutations of the object queries, the loss is defined using the optimal bipartite matching between the predicted and target objects, as computed with an efficient linear assignment problem solver \cite{10.1145/3442348}.
Regression targets are scaled to be of order $\sim 1$ during the loss computation so that they do not dominate the loss or interfere with the matching process.
Finally, the intermediate outputs from each decoder layer are used to compute auxiliary loss terms which are included in the total loss \cite{mask2former}.

\section{Dataset \& Experimental Setup}
\label{sec:experiments}

\subsection{Dataset}\label{sec:dataset}

The TrackML challenge~\cite{2019EPJWC.21406037K,amrouche2020tracking,amrouche2023tracking} was established to promote the development of novel techniques for particle track reconstruction in the high pileup environments anticipated at the HL-LHC.
It has since become a standard benchmark for evaluating track reconstruction methods.
The dataset simulates a generalised LHC-like detector, inspired by planned ATLAS and CMS upgrades for the HL-LHC, and is composed of approximately nine thousand simulated events.
Each event features one hard top quark-antiquark pair ($t\bar{t}$) interaction, overlaid with an additional 200 soft QCD interactions, simulating the expected high pileup conditions at the HL-LHC.

\begin{figure}[htb]
    \centering
    \includegraphics[width=1.0\textwidth]{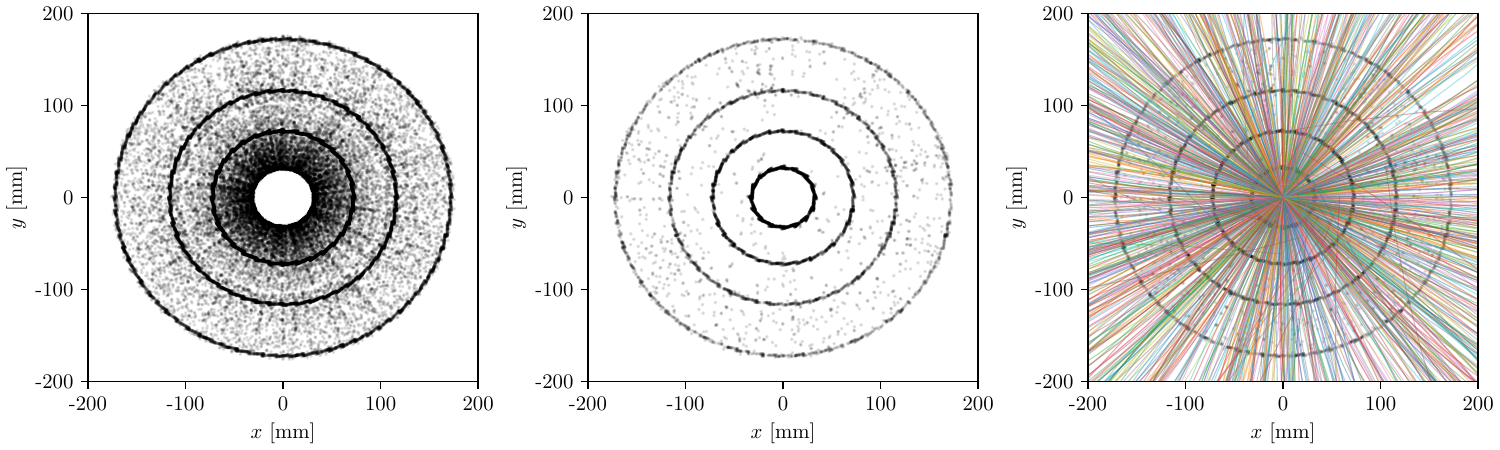}
    \caption{
        Event display in the transverse $x$-$y$ plane, showing a view down the beam-line. This projection captures the cylindrical detector cross-section perpendicular to the beam axis ($z$), with the origin corresponding to the nominal interaction point. Hits and tracks are shown in the transverse plane, where particle trajectories typically curve due to the magnetic field oriented along the $z$-axis.
        (Left) The positions of pixel hits for a single event. (Middle) Hits passing the \SI{750}{\MeV} filter at a cut of 0.1. (Right) Filtered hits along with the trajectories of reconstructable particles (assuming a homogeneous \SI{2}{\tesla} magnetic field) that satisfy $\pt > \SI{1}{\GeV}$ and $|\eta|<2.5$.
    }
    \label{fig:event_display}
\end{figure}

\begin{figure}[htb]
    \centering
    \includegraphics[width=1.0\textwidth]{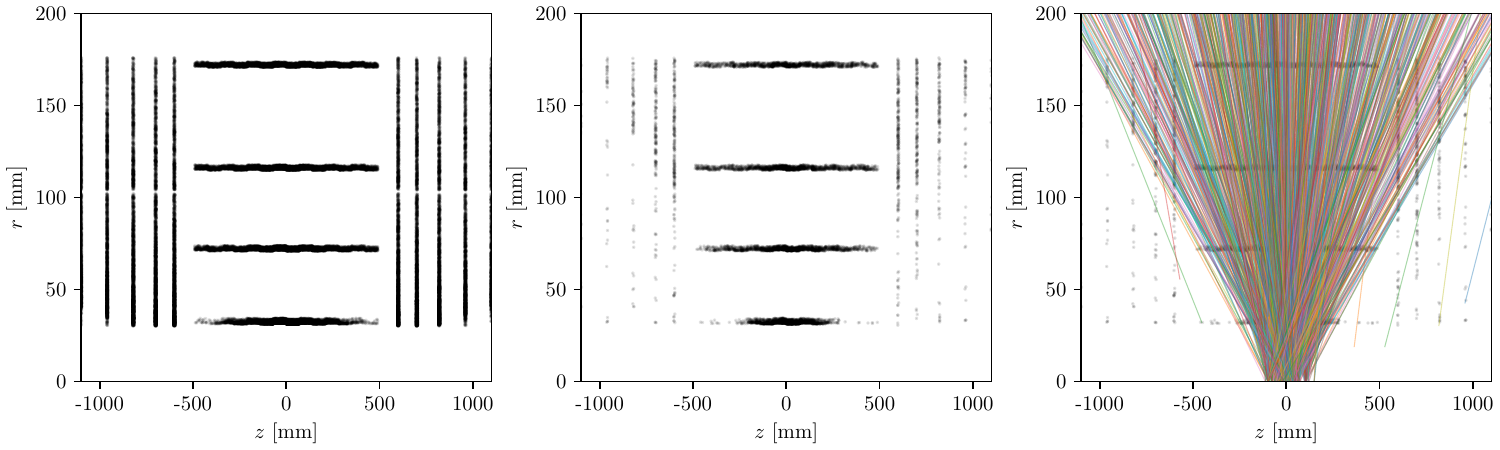}
    \caption{
        Event display in the $r$-$z$ projection of a cylindrical detector. The horizontal axis corresponds to the beam-line direction ($z$), and the vertical axis is the radial distance from the beam-line ($r$). The azimuthal coordinate ($\phi$) is suppressed in this view, such that tracks and hits are projected onto the $r$-$z$ plane irrespective of their azimuthal angle.
        (Left) The positions of pixel hits for a single event. (Middle) Hits passing the \SI{750}{\MeV} filter at a cut of 0.1. (Right) Filtered hits along with the trajectories of reconstructable particles (assuming a homogeneous \SI{2}{\tesla} magnetic field) that satisfy $\pt > \SI{1}{\GeV}$ and $|\eta|<2.5$.
    }
    \label{fig:event_display_rz}
\end{figure}

\begin{figure}[htb]
    \centering
    \includegraphics[width=1.0\textwidth]{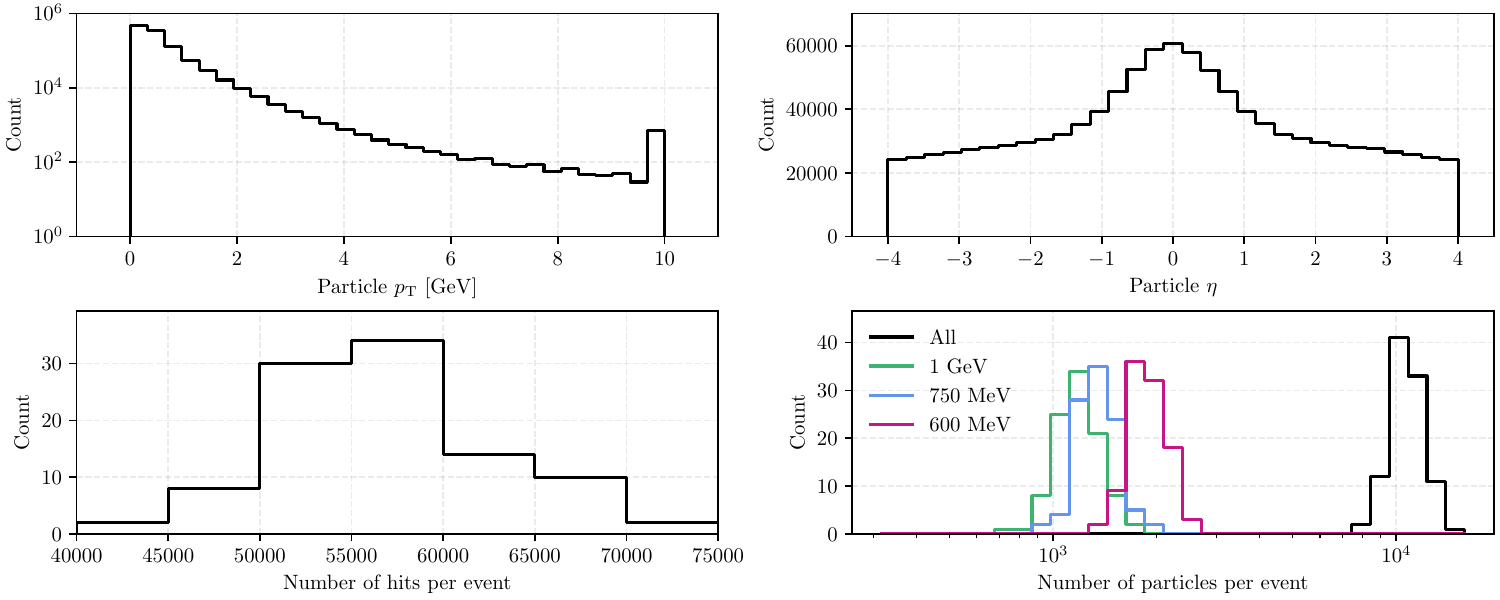}
    \caption{
        Histograms summarising the kinematics and object multiplicities of the TrackML dataset restricted to the inner pixel layers. (Top left) The \pt distribution of all particles with the last bin including overflow. (Top right) A histogram showing the distribution of the pseudorapidity $\eta$ of the particles. (Bottom left) The distribution of the number of pixel layer hits in the events before filtering. (Bottom right) The distribution of the number of total particles in each event. The distribution for all particles is shown in black. Also shown are the number of particles left after applying the three $\pt$ cuts used in this work.
    }
    \label{fig:hists}
\end{figure}

An $x$-$y$ and $r$-$z$ view of a sample event is depicted in \cref{fig:event_display} and \cref{fig:event_display_rz}, respectively.
As shown in \cref{fig:hists}, $\mathcal{O}(10^4)$ particles and $\mathcal{O}(10^5)$ hits are simulated per event, resulting in $\mathcal{O}(10^8)$ tracks to be identified from approximately $\mathcal{O}(10^9)$ hits across the entire dataset.

For each event, TrackML provides information about simulated hits, particles, and their associations.
The detector geometry follows a typical collider layout, with distinct tracking subsystems arranged in concentric layers around the beam axis.
The innermost tracking system consists of silicon pixel sensors arranged in four cylindrical barrel layers, complemented by seven pixel endcap disks on each side.
The outer tracking system comprises strip detectors with six barrel layers and six endcap disks per side.
In this work, we focus exclusively on the innermost pixel detector, including the central barrel and endcap layers.
This restricted geometry may present a more difficult challenge than using all layers, as on average only 4-6 hits are available per track.
However the computational complexity is reduced, making it suitable for our initial studies.

Hits are 3-dimensional spacepoints formed from pixel elements which have been clustered together.
The model is given information about the position of each hit and the shape and orientation of the corresponding clusters.
For the hit position, the superset of the Cartesian and cylindrical coordinates, each defined with the origin at the geometric centre of the detector, is used as this was found to improve convergence during training.
Additional position information is provided in the form of the hit pseudorapidity $\eta$, and the conformal tracking coordinates \cite{hansroul1988fast}.
Finally, information about the associated cluster is provided in the form of the charge fraction (the sum of the charge in the cluster divided by the number of activated channels), the cluster size, and the cluster shape, following the approach in~\cite{fox2021beyond}.

\subsection{Experiments \& Training Setup}\label{sec:train-setup}

In this work, we focus on the task of reconstructing charged particles from hits in the pixel layers of the TrackML detector.
Using only pixel layer information presents a challenging reconstruction task with limited data, requiring the model to efficiently leverage limited information about each particle.
We perform three different training experiments, each with different selection criteria for the reconstructed particles.
In all cases, particles must satisfy the base criterion of having at least three hits left in the pixel layers.
All hits from particles not matching this criteria are labelled as noise and targeted for removal by the hit filtering model.
The transverse momentum and pseudorapidity selections vary across the three experimental configurations:
\begin{enumerate}
    \setlength{\itemsep}{0pt}
    \item \ptmin = \SI{600}{\MeV}: Pseudorapidity selection of $\etamax = 2.5$
    \item \ptmin = \SI{750}{\MeV}: Pseudorapidity selection of $\etamax = 2.5$
    \item \ptmin = \SI{1}{\GeV}: Wider pseudorapidity selection of $\etamax = 4.0$, demonstrating the model's capability to reconstruct tracks efficiently across the full detector acceptance
\end{enumerate}
For each experiment, a hit filtering model is first trained to select hits from reconstructable particles.
A tracking model is trained to reconstruct tracks based on the hits that pass the filtering model's selection.
The filtering models each have approximately 8M trainable parameters, while the tracking models have approximately 22M trainable parameters.
A summary of the different experiments is provided in \cref{tab:dataset}.
The different choices of \ptmin and \etamax allow us to explore the trade-offs between model complexity, inference time, and performance.
They also provide an insight into the impact of reducing the effective training statistics (i.e. the number of target particles), as discussed in \cref{sec:results}.

For each of the validation and testing sets, 100 events are reserved
with the remaining events used for training.
Models are trained on a single NVIDIA A100 GPU for 30 epochs.
The hit filtering models trained in approximately 10 hours, while the tracking models trained in 20--60 hours, depending on the \ptmin and \etamax selections applied.
A batch size of a single event is used.
Inference times are provided in \cref{sec:timing}.

\begin{table}[H]
    \centering
    \begin{tabular}{cccccc}
        \toprule
        \midrule
        \ptmin & \etamax & Hits (Pre) & Hits (Post) & Particles & Object Queries\\
        \midrule
        \SI{600}{\MeV} & 2.5 & 57k & 12k & 1800 & 2100 \\
        \SI{750}{\MeV} & 2.5 & 57k & 8k & 1300 & 1800 \\
        \SI{1}{\GeV}   & 4.0 & 57k & 11k & 1100 & 1500 \\
        \midrule
        \bottomrule
    \end{tabular}
    \vspace{1em}
    \caption{
        Summary of the models trained with different \ptmin and \etamax selections.
        For each selection, the number of hits pre- and post-filtering is shown, along with the average number of target particles in the event, and the configured number of object queries for the associated track reconstruction model.
        Hit and particle counts are averaged over the test set.
    }
    \label{tab:dataset}
\end{table}

\section{Results}\label{sec:results}

As described in \cref{sec:dataset}, we consider a particle to be reconstructable if it leaves at least three hits in the pixel detector, has an absolute pseudorapidity less than \etamax, and a transverse momentum $\pt > \ptmin$.
The hit filtering performance is discussed in \cref{sec:hit-filtering}, while the tracking performance is discussed in \cref{sec:tracking} and inference times are presented in \cref{sec:timing}.
All results shown are obtained using the test set of 100 events.

\subsection{Hit Filtering}\label{sec:hit-filtering}

In the hit filtering task, hits are labelled as \textit{signal} if they belong to a reconstructable particle and \textit{noise} otherwise.
The filter performance is evaluated using binary efficiency and purity.
The hit efficiency is defined as the fraction of signal hits retained after filtering, while the purity represents the fraction of retained hits that are signal hits.

\begin{figure}[htb]
    \centering
    \includegraphics[width=\textwidth]{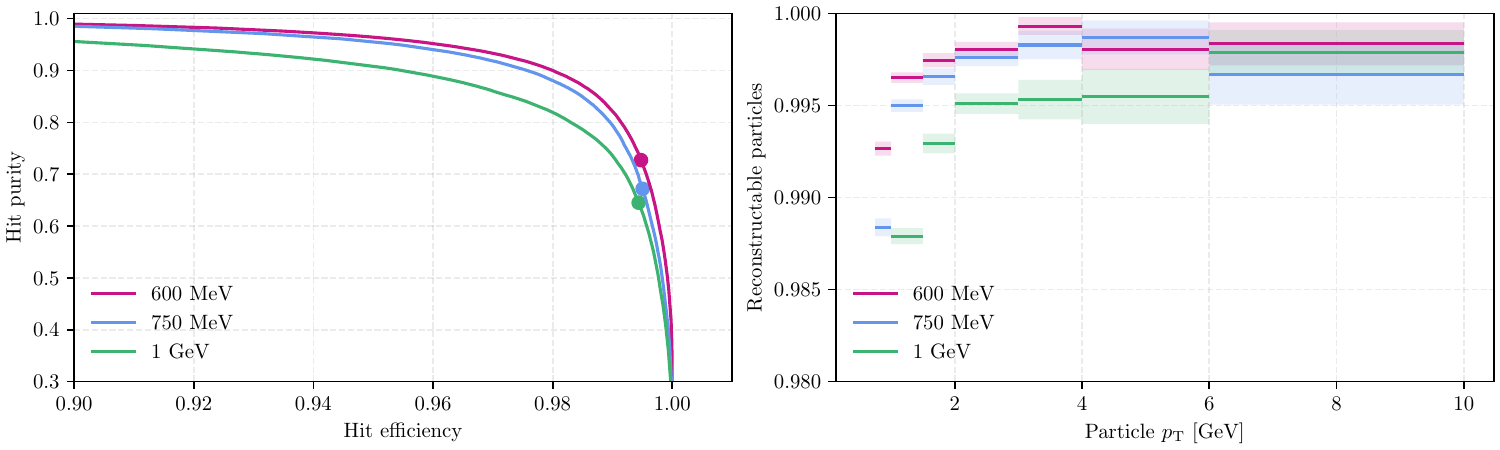}
    \caption{
        Hit filtering performance for each of the three models.
        (Left) Signal hit purity as a function of the signal hit efficiency.
        The markers show the efficiency and purity at the chosen threshold of 0.1 and each model achieves an area under the curve of 0.998.
        (Right) The fraction of particles that remain reconstructable as a function of simulated particle \pt after filtering hits that fall below the 0.1 threshold.
        Binominal errors are indicated by the shaded regions, and the final bin includes overflow.
    }
    \label{fig:filter_response}
\end{figure}

\cref{fig:filter_response} demonstrates how these metrics vary with the probability threshold used to classify hits.
At our chosen operating threshold of 0.1, the models achieve impressive performance while significantly reducing the input multiplicity for downstream tracking.
The dramatic reduction in hit multiplicity achieved, from approximately 57k to 6k-12k hits depending on the model, are detailed in \cref{tab:dataset} and also visualised in \cref{fig:event_display}.

As summarised in \cref{tab:filtering-perf}, the \SI{600}{\MeV} model maintains a hit efficiency of 99.5\% while improving purity from 15.6\% pre-filter to 72.7\% post-filter.
Similarly, the \SI{750}{\MeV} and \SI{1}{\GeV} models achieve efficiencies of 99.6\% and 98.8\% respectively, with corresponding purities of 67.2\% and 64.5\%.
This represents dramatic purity improvements from their pre-filter values of 11.2\% and 13.1\%.
When considering only hits from particles passing the requirement on $\etamax$ (and excluding noise particles), the initial purities are higher at 34.0\%, 24.3\%, and 14.8\% respectively, however the post filter purity still represents a significant improvement.

\begin{table}[htb]
    \centering
    \begin{tabular}{lcccc}
        \toprule
        \midrule
        Model  & Initial Purity ($|\eta| < \etamax$) & Filter Efficiency & Filter Purity & Reconstructable \\
        \midrule
        \SI{600}{\MeV} & 15.6\% (34.0\%)  & 99.5\% & 72.7\% & 99.7\% \\
        \SI{750}{\MeV} & 11.2\% (24.3\%) & 99.5\% & 67.2\%  & 99.6\% \\
        \SI{1}{\GeV}   & 13.1\% (14.8\%) & 99.4\% & 64.5\% & 98.8\% \\
        \midrule
        \bottomrule
        \vspace{0.0cm}
    \end{tabular}
    \caption{
        Performance of the hit filtering models on the test set.
        From left to right the columns show the initial hit purity (and the purity calculated using only hits from particles that fall within the specified \etamax excluding detector noise hits not associated with any particle), the post-filter efficiency and purity, and the percentage of particles which remain reconstructable (i.e. retain three or more hits in the pixel layers) after the application of the filter.
        \etamax is 2.5 for the \SI{600}{\MeV} and \SI{750}{\MeV} models and 4 for the \SI{1}{\GeV} model.
        Statistical uncertainties are negligible.
    }
\label{tab:filtering-perf}
\end{table}

We also examine the fraction of particles that remain reconstructable after filtering (i.e. those that retain at least three pixel hits).
The results are shown as a function of the particle \pt in \cref{fig:filter_response}, and integrated values are shown in \cref{tab:filtering-perf}.
For particles with $\pt > \SI{1}{\GeV}$, each model achieves excellent performance, preserving more than 99.5\% of particles.
The performance for each model degrades as the particle \pt approaches the \ptmin used to train the model, indicating that the \ptmin threshold needs to be set below the desired minimum particle \pt for reconstruction.
Given the strong performance of the hit filtering model, we anticipate that this approach may find use as a pre-burner to other track reconstruction algorithms, including traditional approaches.

\subsection{Tracking}\label{sec:tracking}

Track reconstruction performance is evaluated in terms of the efficiency and fake rate. 
Tracking efficiency is defined as the fraction of reconstructable particles that are correctly matched to a reconstructed track.
The fake rate is defined as the fraction of reconstructed tracks that are not well-matched to any reconstructable particle.
These tracks either result from accidental combinations of unrelated hits, or fail to meet the matching threshold due to relevant hits not being included in the predicted track mask or insufficient detector activity from the corresponding particle to have enough hits to satisfy the matching criteria. 
A lower fake rate indicates better suppression of such spurious predictions.
We consider two different criteria to define whether a track is matched to a particle or not: \textit{double majority} and \textit{perfect} matching.
Under the double majority (DM) criteria, a match occurs if $>\pct{50}$ of the particle's hits are assigned to the track and $>\pct{50}$ of the hits on the track are from that particle.
Under the perfect criteria, a match occurs if all of the particle's hits are assigned to the track, and no other hits are assigned.
Each track is matched to the simulated particle that contributes the largest number of hits to the track.
If two particles have the same number of hits on a given track, one is chosen at random.
The efficiencies using the DM and perfect match criteria are noted as $\effdm$ and $\effperfect$, respectively.
Subscripts are used to indicate the minimum $\pt$ of the matched simulated particles included in the calculation.
The fake rate \fakedm~is defined using the DM matching and without any selections on matched simulated \pt.
To enable comparison with other methods, we also define a fake rate \fakedmpt, which considers only tracks matched to target particles with $\pt > \SI{0.9}{\GeV}$.
The particle-level inefficiencies resulting from the filtering step are included in the tracking efficiencies discussed here.
The duplicate rate $d$ is defined as the fraction of reconstructed tracks that have identical hit assignments.

\begin{figure}[htbp]
    \centering
    \includegraphics[width=1.0\textwidth]{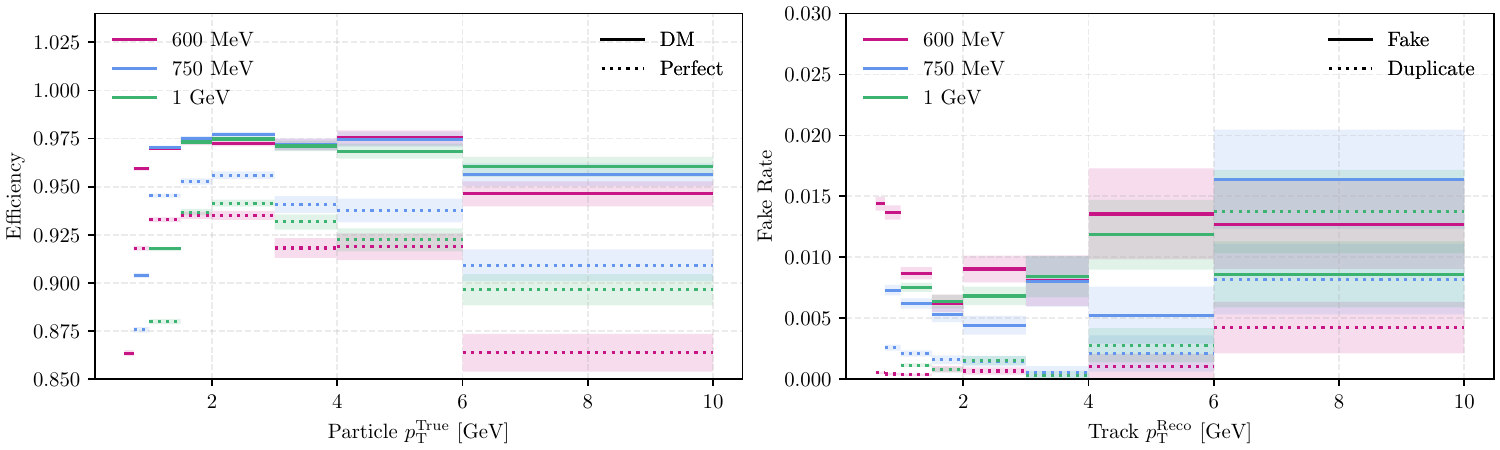}
    \caption{
        (left) Track reconstruction efficiency as a function of the simulated particle \pt for the DM (solid) and perfect (dashed) matching criteria.
        (right) Fake rate using the DM matching criteria (solid) and duplicate rate (dashed) as a function of the reconstructed track \pt.
        The shaded regions indicate the binomial errors, and the final bin includes overflow.
    }
    \label{fig:pt_eff_fr}
\end{figure}
\begin{figure}[htbp]
    \centering
    \includegraphics[width=1.0\textwidth]{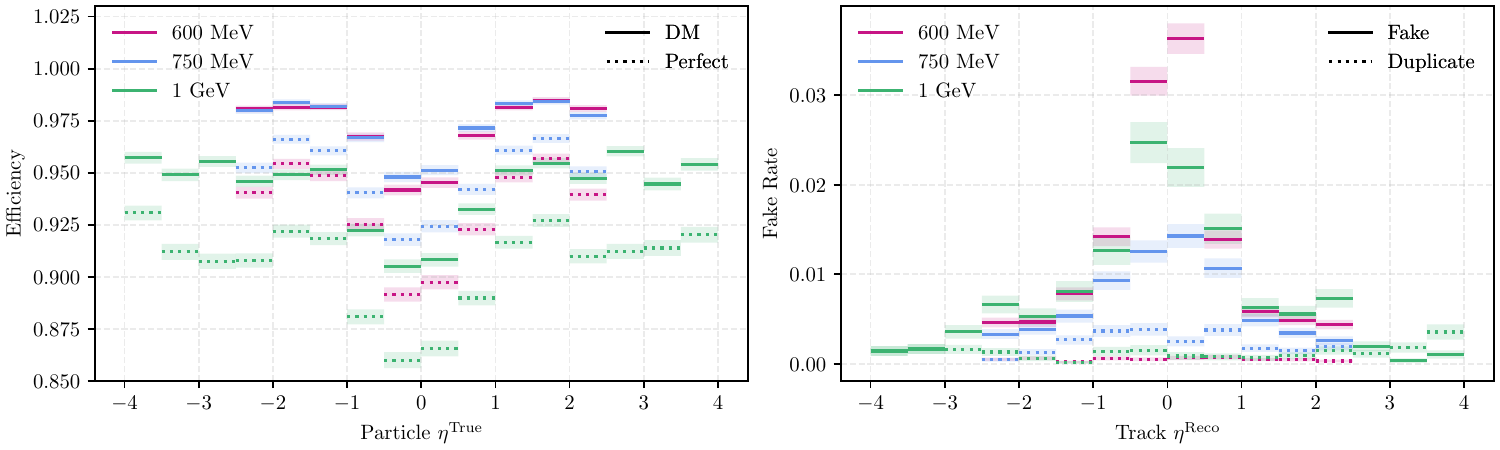}
    \caption{
        (left) Track reconstruction efficiency as a function of the simulated particle $\eta$ for the DM (solid) and perfect (dashed) matching criteria.
        For consistency, only particles with $\pt > \SI{1}{\GeV}$ are considered.
        (right) Fake rate using the DM matching criteria (solid) and duplicate rate (dashed) as a function of the reconstructed track $\eta$.
        The shaded regions indicate the binomial errors.
    }
    \label{fig:eta_eff_fr}
\end{figure}

The tracking performance of the different models depends on the values of \ptmin and \etamax used to define the target particles during training.
As shown in \cref{fig:pt_eff_fr}, all three models achieve a plateau efficiency of approximately 97.5\% for particiles satisfying $2 < \pt < 6$~GeV.
Below \SI{2}{\GeV}, the efficiency drops significantly as the particle \pt approaches the \ptmin used to train the model, and falls to nearly zero below this threshold.
The reconstruction efficiency for particles with $\pt > \SI{6}{\GeV}$ drops slightly from the plateau to around 95--96\% depending on the model, however the statistical uncertainties are large in this region due to the sharp drop decrease in particle multiplicity with increasing \pt.
Notably, the models are able to perfectly reconstruct particles at high rates (above \pct{90} in most cases).
However, a stronger decreasing trend with \pt is observed for the perfect match, reflecting the increased difficulty in assigning hits at high \pt.
The overall high performance underscores the models ability to handle the combinatorial complexity of the TrackML dataset, with only a low number of output tracks not perfectly reconstructed (around 2-4\% depending on the model and \pt).

\cref{fig:pt_eff_fr} also shows the fake and duplicate rates as a function of the reconstructed track \pt.
In general, the fake rate remains low but shows an increasing trend with \pt. 
The \SI{600}{\MeV} model has a significantly higher fake rate at low \pt compared to the \SI{750}{\MeV} model (\pct{1.5} compared to \pct{0.7}), possibly due to the increased complexity associated of reconstructing low \pt particles.
The rate of duplicate tracks (those with identical hit assignments) is also examined, and is comfortably below \pct{0.5} for tracks below \SI{6}{\GeV} for all models.
Above \SI{6}{\GeV}, the duplicate rate increases to 0.5--1.5\% depending on the model, although statistical uncertainties are large.

The efficiencies, fake rates, and duplicate rates are shown as a function of $\eta$ in \cref{fig:eta_eff_fr}.
The efficiency remains high across the full $\eta$ range, with a slight decrease of approximately \pct{2.5} in the central region ($|\eta| < 1$) for all models.
A corresponding increase in the fake rate is observed in this region, consistent with the increased particle density and \pt in this region.
The efficiency of the \SI{1}{\GeV} model is slightly lower than the other two models due to the drop in reconstruction efficiency for particles with \pt approraching \ptmin, as discussed above.
The duplicate rate remains low across the full $\eta$ range, with a slight increase in the central region.

Performance comparisons with existing methods are summarised in \cref{tab:tracking_comparison}, though several important methodological differences should be noted.
While OC and HGNN include particles in the forward region ($|\eta| < 4$), our \SI{600}{\MeV} and \SI{750}{\MeV} models focus on the more challenging central region ($|\eta| < 2.5$) where track density is highest, and our \SI{1}{\GeV} model uses $|\eta| < 4$.
Given the reduced performance near the \pt threshold, the \SI{750}{\MeV} model provides the most meaningful comparison with OC.
Despite the expectation that particles in the forward region are easier to reconstruct due to lower track density, the \SI{750}{\MeV} model achieves a slightly higher efficiency ($\effdm=97.2\%$) than OC ($96.4\%$) while reducing the fake rate by a factor of three to $\fakedmpt=0.3\%$.
In addition, our approach demonstrates improved perfect match efficiency, increasing the rate by 9 percentage points.
The HGNN method includes hits on the strip layers and requires more than 5 hits for a particle to be reconstructable, and would require additional post-processing steps after the model to reduce the high fake rate.
While these differences complicate direct comparisons, our results achieve comparable efficiencies with significantly better fake rate, all whilst using less information and being significantly faster as discussed in \cref{sec:timing}.

\begin{table}[htbp]
    \centering
    \begin{tabular}{lcccccccc}
    \toprule
    \midrule
    & \effperfect & \effperfectn & \effdm & \effdmn & \fakedm & \fakedmpt & $d$               \\
    \midrule
    \SI{1}{\GeV}                & 90.4\% & -      & 94.1\% & -      & 0.7\%  & -     & 0.1\%  \\ 
    \SI{750}{\MeV}              & 94.8\% & 94.5\% & 97.2\% & 97.1\% & 0.7\%  & 0.3\% & 0.2\%  \\
    \SI{600}{\MeV}              & 93.2\% & 93.0\% & 97.1\% & 97.0\% & 1.2\%  & 0.2\% & 0.1\%  \\
    \midrule
    OC \cite{lieret2024object}  & -      & 85.8\% & -      & 96.4\% & -      & 0.9\% & -      \\
    HGNN \cite{trackhgnn}       & -      & -      & 97.9\% & -      & 36.7\% & -     & -      \\
    \midrule
    \bottomrule
    \vspace{0.0cm}
    \end{tabular}
    \caption{
        Comparison of the results for the various models using the TrackML dataset on the test set.
        For the efficiencies, two different \pt thresholds are shown to facilitate comparisons with existing work.
        OC \cite{lieret2024object} uses $\ptmin = \SI{0.9}{\GeV}$, while HGNN \cite{trackhgnn} uses $\ptmin = \SI{1}{\GeV}$ and also includes hits from the strip layers.
        Both OC and HGNN attempt to reconstruct particles satisfying $|\eta| < 4$, whereas we target $|\eta| < 2.5$ for the \SI{600}{\MeV} and \SI{750}{\MeV} models and $|\eta| < 4.0$ for the \SI{1}{\GeV} model.
        Statistical uncertainties are negligible.
    }
    \label{tab:tracking_comparison}
\end{table}

\begin{figure}[htbp]
    \centering
    \includegraphics[width=1.0\textwidth]{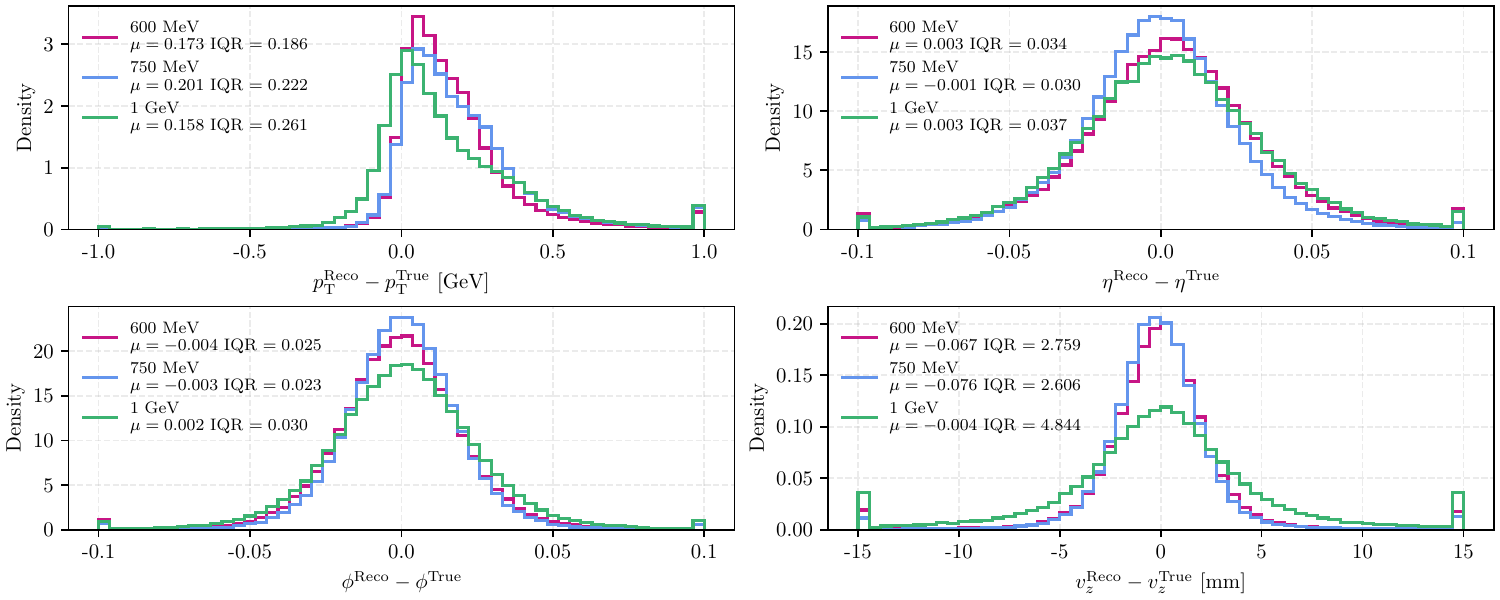}
    \caption{
        Residuals between the regressed parameters of reconstructed tracks and the targets from the DM matched tracks.
        Tracks matched to simulated particles with $\pt > \ptmin$ are used for each model.
        The mean $\mu$ and inter-quartile range (IQR) of the residuals for each model is shown in the legend.
        Residuals that lie outside the plot range are placed into extremal bins.
    }
    \label{fig:regr_residual}
\end{figure}

\cref{fig:regr_residual} shows the residuals between reconstructed and target track parameters, where the target parameters are derived from DM matched particles.
While $v_z$ is regressed directly, the \pt, $\eta$, and $\phi$ are constructed from the track's Cartesian momentum.
The residuals demonstrate minimal bias, with the exception of the \pt residuals which have a positive skew, possibly due to boundary effects or the relative abundance of \lowpt particles~\cite{cao2024systematic}.
However, the current regression performance does not yet match the precision achieved by established experiments like ATLAS and CMS, which employ highly tuned parametric fitting techniques~\cite{IDTR-2022-04, CMS-TRK-11-001}.
Regardless, these results serve as a promising proof-of-principle for our novel approach of jointly performing hit-to-track assignment and track parameter fitting in a single model, and may already be sufficiently precise for simple triggering applications.
Several avenues exist for improving the parameter estimation.
First, incorporating hits from the outer strip layers would significantly enhance momentum resolution.
Secondly, directly regressing the parameters of interest, rather than constructing them from the Cartesian momentum, would likely further improve performance.
Finally, the model could be extended to estimate track parameter uncertainties and their correlations, bringing the approach closer to the comprehensive track fitting methods used in production environments.
Alternatively, a KF or least-squares fit could be applied to the set of hits assigned to a given track prediction, as is done in ATLAS and CMS, to extract high-precision track parameters.

Analysis of model performance versus training set size for the \SI{1}{\GeV} model reveals that the model has not yet reached performance saturation when using the complete TrackML dataset.
This suggests significant potential for improved performance through additional training data. 
Furthermore, our approach provides a convenient trade-off between model size, performance, and inference speed: when inference speed is critical, model size can be reduced to achieve faster processing times, while in applications where speed is not the primary concern, model size can be increased to enhance tracking performance.
This flexibility allows the same fundamental approach to be adapted to different experimental requirements and computational constraints.

\subsection{Inference Time}
\label{sec:timing}

Inference time is a critical metric for track reconstruction at particle colliders, directly impacting the feasibility of real-time event processing. Our Transformer-based models demonstrate competitive performance in this key metric. The hit filtering forward pass requires on average approximately $23 \pm \SI{3}{\ms}$ to process a single event comprising O(60k) hits when using an NVIDIA A100 GPU.
The hit filtering model is Just-In-Time (JIT)-compiled via \texttt{torch.compile}, which significantly accelerates the forward pass during inference.

The tracking inference time depends on the choice of $\ptmin$, and ranges from approximately \SI{75}{\ms} for the \SI{750}{\MeV} and \SI{1}{\GeV} models to \SI{100}{\ms} for the \SI{600}{\MeV} model, as detailed in \cref{tab:inference_timing}. 
For the \SI{750}{\MeV} model, which represents a good balance between tracking performance and inference time, the total time for hit filtering and tracking combined is approximately \SI{97}{\ms} on average.

\begin{table}[htb]
    \centering
    \begin{tabular}{lcc}
    \toprule
    \midrule
    & Tracking time [ms] & Filter + tracking time [ms] \\
    \midrule
    \SI{600}{\MeV} & $100 \pm 12$ & $123 \pm 14$ \\
    \SI{750}{\MeV} & $74 \pm 9$  & $97 \pm 11$ \\
    \SI{1}{\GeV}   & $76 \pm 9$  & $99 \pm 11$  \\
    \midrule
    \bottomrule
    \vspace{0.0cm}
    \end{tabular}
    \caption{
        Mean inference times and standard deviations for the tracking model forward pass, and combined filtering and tracking forward pass, evaluated on an NVIDIA A100 GPU with a batch size of 1.
    }
    \label{tab:inference_timing}
    \vspace{-0.5cm}
\end{table}

\begin{figure}[htbp]
    \centering
    \includegraphics[width=0.9\textwidth]{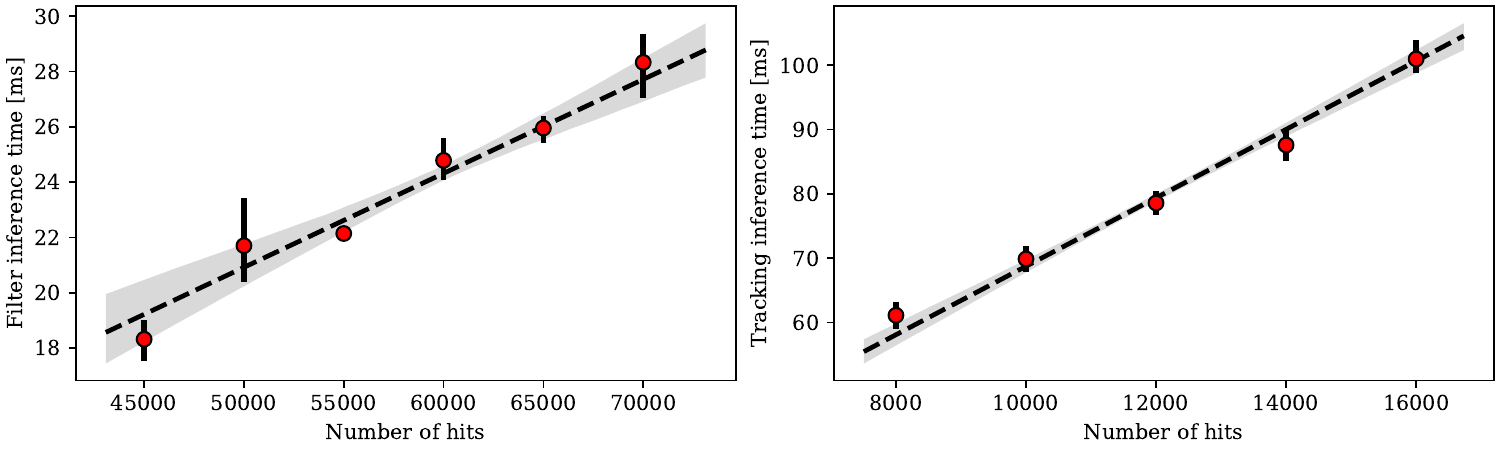}
    \caption{
        Inference times as a function of the input hit multiplicity for the hit filtering model (left) and \SI{1}{\GeV} track reconstruction model (right), evaluated on an NVIDIA A100 GPU with a batch size of 1.
        The average time in each bin and the \pct{95} confidence intervals are shown by the markers and vertical error bars.
        A linear fit is shown in the dashed line, along with a \pct{95} confidence interval shown by the shaded region.
    }
    \label{fig:inference_timing}
\end{figure}

To understand scaling behavior in order to estimate the timing requirements to consider all detector layers, we analysed inference times as a function of hit multiplicity in \cref{fig:inference_timing}.
Both the filtering and tracking models exhibit linear scaling with the number of input hits, a critical result achieved through the architectural choices described in \cref{sec:Transformers}. 
This linear scaling is crucial for adaptability to varying event complexities.

The linear scaling observed in \cref{fig:inference_timing} enables us to extrapolate performance to the higher hit multiplicities anticipated in the full detector of a typical HL-LHC event.
For the $\mathcal{O}(350\mathrm{k})$ hits expected in the ITk detector at the start of Run 4~\cite{itk_tdr}, we project a combined filter and tracking time of approximately $\SI{800}{\ms}$ to reconstruct tracks with $\ptmin = \SI{1}{\GeV}$ and $\etamax = 4$, assuming this linear scaling holds.

\begin{figure}[htbp]
    \centering
    \includegraphics[width=0.6\textwidth]{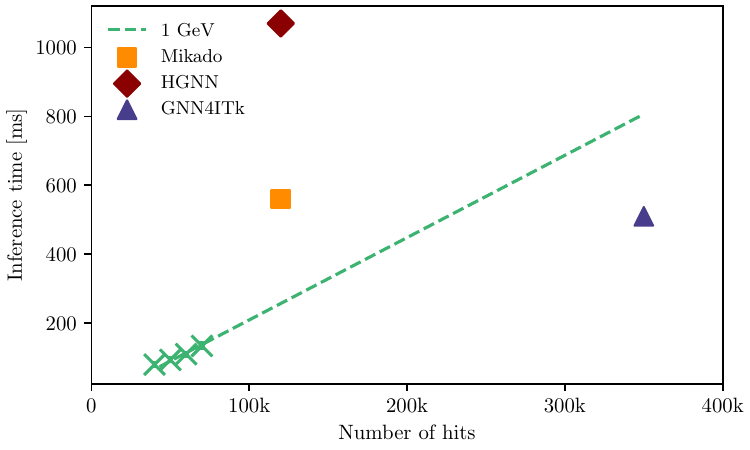}
    \caption{
        Combined filtering and tracking inference times for the MaskFormer track reconstruction model ($\SI{1}{\GeV}$, trained with $\etamax = 4$) as a function of input hit multiplicity.
        A linear fit to the MaskFormer data is shown and, assuming its validity, extrapolated to the O(350k) hits expected in the full ITk detector.
        This extrapolation allows for an approximate comparison with other approaches: Mikado~\cite{amrouche2023tracking}, HGNN~\cite{trackhgnn}, and GNN4ITk~\cite{ATL-PHYS-PUB-2024-018}.
        Direct comparisons should be approached with caution due to significant methodological differences: MaskFormer, Mikado, and HGNN are evaluated on the TrackML dataset, while GNN4ITk uses a realistic ITk detector simulation. 
        Furthermore, MaskFormer, HGNN, and GNN4ITk are benchmarked on NVIDIA A100 GPUs, whereas Mikado is evaluated on 2 CPU cores. 
        Additional variations in detector geometries, event multiplicities, and target particle definitions may also affect comparisons.
        This figure shows that the model presented has great potential to be the state-of-the-art solution for a highly combinatoric process: fast and accurate charged particle reconstruction.
    }
    \label{fig:inference_timing_extrap}
\end{figure}

Direct comparisons of inference times across different track reconstruction methods are inherently challenging due to variations in experimental setups and differences in reported metrics.
These include differences in detector geometry, the total number of hits considered, the inclusion of various subdetectors, and the choices of target particle definitions.

In light of these caveats, our projected inference time of $\sim \SI{800}{\ms}$ for the full ITk detector is comparable to recent, optimised results from the GNN4ITk project~\cite{ATL-PHYS-PUB-2024-018}, which represents a highly customised approach developed through extensive, multi-year optimisation efforts.
Remarkably, our end-to-end learned approach achieves comparable timing performance while leveraging largely off-the-shelf ML architectures demonstrating a strategy that offers potential advantages in flexibility and broader applicability, despite requiring significantly less development effort.
Furthermore our results improve upon the timing studies in \rcite{trackhgnn} and are comparable with optimised non-ML approach to trigger-level tracking~\cite{atlas_phase2_trig}.

In this study, MaskFormer achieves a high efficiency and low fake rate using exclusively pixel detector hits.
While incorporating strip hits could further improve performance, it would also increase hit multiplicity and thus processing time.
A key finding is the strong tracking performance achieved without this additional information, suggesting a potential advantage: for certain applications, the computational cost of processing all strip hits might be avoidable if pixel-based performance is sufficient.
Thus, the $\approx \SI{800}{\ms}$ projection for all $\mathcal{O}(350\mathrm{k})$ ITk hits might be an upper limit if strong performance is maintained by primarily using pixel data or a subset of strip hits, enabling faster inference.

The timing estimates assume continued linear scaling with hit multiplicity. In realistic detector environments additional factors, such as geometric irregularities or non-uniform magnetic fields, may necessitate larger $\phi$ windows and introduce deviations from this behaviour. Hardware constraints, including the inability to fit the full model and input data into device memory, could render training infeasible. Validating these extrapolations using full detector simulations remains an important future step.

Nevertheless, the current inference times have been achieved with minimal optimisation, indicating substantial potential for future speed improvements.
Our approach is poised for significant enhancements through architecture optimisation, enabling JIT compilation for the full tracking model, and the application of established techniques such as model pruning and quantisation.
Additional gains are anticipated from training smaller, more efficient models on expanded datasets, while inference throughput can be further boosted by increasing the batch size.
These clearly defined avenues for development are expected to markedly enhance the real-world applicability and performance of our method.

\FloatBarrier

\section{Conclusion}\label{sec:conclusion}

We present a novel application of the Transformer architecture to charged particle track reconstruction, achieving state-of-the-art results on the full TrackML dataset.
Our approach demonstrates strong performance, with the \SI{750}{\MeV} model achieving an efficiency for particles with $\pt > \SI{1}{\GeV}$ of approximately 97.2\% and a fake rate of 0.7\%.
We also demonstrate the model's capacity to reconstruct low-\pt tracks with the \SI{600}{\MeV} configuration, and tracks across the full $|\eta|$ range with the \SI{1}{\GeV}, $\etamax = 4$ model.
The model's inference time of approximately \SI{100}{\ms} makes it highly competitive for real-world applications.

This success stems from two key innovations: a Transformer-based hit filtering stage that effectively reduces input multiplicities without requiring complex graph construction, and a MaskFormer track reconstruction stage that leverages cross-attention to build an explicit latent representation of each track.
This unified approach enables simultaneous track finding and parameter estimation while naturally handling hit sharing between tracks.
The model's linear scaling with hit multiplicity, enabled by efficient attention kernels and demonstrated in our timing studies, suggests promising scalability for future detector environments.
By aligning with recent advancements in the field of machine learning, our approach stands to benefit from future developments, offering scalable, adaptable, and high-performance solutions to meet the increasing computational demands of high-energy physics.

While our method demonstrates promising performance on the TrackML dataset, we note that real detector environments introduce additional complexities not included in this dataset.
For example, including non-uniform geometries, material interactions, sensor inefficiencies, and magnetic field inhomogeneities can all affect performance and time-scaling behaviours.
Evaluating the robustness of Transformer-based tracking in such settings remains an important future direction. 
It will also be critical to assess generalisability and deployment feasibility by testing on simulated event in the planned ATLAS and CMS tracking detector upgrades.

There are several directions for future work to enhance the capabilities and applicability of the model.
First, we plan to improve the algorithmic efficiency of the model, allowing it to handle the increased hit multiplicities that would stem from including the strip hits, and reconstructing additional particles with even lower transverse momenta.
Second, combining the hit filtering and track reconstruction stages into a single model could streamline the training process and reduce inference times.
Third, we plan to test the model and scaling properties on more realistic detector simulations, including irregularities in the detector layout, material effects, and magnetic field variations to validate generalisation.
Finally, incorporating additional information from other detector systems, such as calorimeters, could further enhance the reconstruction of charged particles and enable the simultaneous reconstruction of charged and neutral particles, paving the way toward full particle-flow–style reconstruction.

In the near term, the architecture is already suitable for hybrid use within existing reconstruction pipelines. 
The model's tunable nature, enabling trade-offs between performance and inference time and targeting of specific particle kinematics, makes it adaptable to various high-energy physics experiments, from trigger-level systems to offline reconstruction, without requiring changes to established reconstruction techniques.
The hit filtering component can be readily integrated as a fast pre-processing step to reduce input multiplicity.
Similarly, the tracking model could act as a learned track-seeding mechanism, replacing or augmenting standard pattern recognition algorithms and integrating seamlessly with current pipelines to deliver high-quality track parameter estimates.

Beyond these immediate uses, our work contributes to a broader shift in reconstruction strategy. An end-to-end optimisable framework for charged particle reconstruction, such as the one presented, could offer significant benefit to a diverse range of experimental setups beyond traditional colliders experiments.
Most significantly, the success of Transformer-based architectures in both tracking and vertexing points toward a unified approach to particle physics reconstruction, moving away from specialised solutions and toward generalised learned models leveraging cross-domain advances in machine learning. Such models could significantly impact how we process and analyse particle collision data and meet the evolving challenges of high-energy physics.

\section*{Acknowledgments}

We gratefully acknowledge the support of the UK's Science and Technology Facilities Council (STFC). S.V.S is supported by ST/X005992/1. P.D., M.H., and N.P. are supported by the STFC UCL Centre for Doctoral Training in Data Intensive Science  (ST/W00674X/1) and by departmental and industry contributions. G.F and T.S. receives support from the STFC (ST/W00058X/1) and T.S. is supported by the Royal Society (URF/R/180008). S.R. is supported by CERN, the Banting Postdoctoral Fellowship program, and the Natural Sciences and Engineering Research Council of Canada. We also extend our thanks to UCL for the use of their high-performance computing facilities, with special thanks to Edward Edmondson for his expert management and technical support.

\printbibliography
\end{document}